\documentclass{ar2e}
\newcommand{\beqn}{\begin{eqnarray}}
\newcommand{\beq}{\begin{equation}}
\newcommand{\eeqn}{\end{eqnarray}}
\newcommand{\eeq}{\end{equation}}

\newcommand{\Frac}[2]{\frac{\displaystyle{#1}}{\displaystyle{#2}}}
\newcommand{\lsim}{\raise0.3ex\hbox{$\;<$\kern-0.75em\raise-1.1ex\hbox{$\sim\;$}}}
\newcommand{\gsim}{\raise0.3ex\hbox{$\;>$\kern-0.75em\raise-1.1ex\hbox{$\sim\;$}}}

\begin{document}
\input epsf.def   
\input epsf
\title{NEW PHYSICS IN $CP$ VIOLATION EXPERIMENTS}

\markboth{MASIERO \& VIVES}{NEW PHYSICS IN $CP$ VIOLATION EXPERIMENTS}

\author{Antonio Masiero
\affiliation{SISSA, Via Beirut 2--4, 34013 Trieste, Italy and INFN, sez. di 
Trieste, Trieste, Italy}
Oscar Vives
\affiliation{Departament de F\'{\i}sica Te\`orica and IFIC,
Universitat de Val\`encia-CSIC, E-46100, Burjassot (Val\`encia), Spain}}


\begin{abstract}
$CP$ violation plays a privileged role in our quest for new physics
beyond the electroweak standard model (SM). In the SM the violation of
$CP$ in the weak interactions has a single source: the phase of the quark
mixing matrix (the CKM matrix, for Cabibbo--Kobayashi--Maskawa).  
Most extensions of the SM exhibit new sources of
$CP$ violation. For instance, the truly minimal supersymmetric extension
of the SM (CMSSM) has two new phases in addition to the CKM phase.
Given that $CP$ violation is so tiny in the kaon system, is still
largely unexplored in $B$ physics and is negligibly small 
in the electric dipole moments, it is clear that
new physics may have a good chance to manifest some departure from the
SM in this particularly challenging class of rare phenomena. On the
other hand, it is also apparent that $CP$ violation generally represents
a major constraint on any attempt at model building beyond the SM. In
this review we tackle these two sides of the relation between $CP$
violation and new physics. Our focus will be on the potentialities
to use $CP$ violation as a probe on Supersymmetric (SUSY) extensions of 
the SM. We wish to clarify the extent to which such indirect signals for SUSY
are linked to a fundamental theoretical issue: is there a relation
between the mechanism that originates the whole flavor structure and
the mechanism that is responsible for the breaking of supersymmetry?
Different ways to answer this question lead to quite different
expectations for $CP$ violation in B physics.
\end{abstract}

\maketitle

\section{INTRODUCTION}

With the shutdown of the Large Electron--Positron Collider (LEP) at CERN, a 
crucial and glorious chapter of twentieth century physics has ended. 
Thanks to the efforts of experimentalists and theorists all over the world,
we are now able to make a definitive statement concerning our 
knowledge of the physical world: the Standard Model (SM) represents the 
correct description of fundamental interactions up to  energies of ${\cal O}
(100\  \mbox{GeV})$. The fact 
that we know the right theory of fundamental interaction working at 
distances as small as $10^{-18}$ m is a tremendous achievement. 

Having emphasized this bright side of the last two decades, 
one has must also admit that the experimental and theoretical particle 
physics of 
these years has been rather unsuccessful in finding a road to follow beyond 
the SM. If we are all convinced that the SM 
cannot be the ultimate theory of ``everything'' (if only 
because it does not include gravity), we are in the dark about 
when, where and how new physics beyond the SM should manifest 
itself. Searches for new particles and interactions in leptonic and 
hadronic machines have failed to produce results. Theoretically, the most 
promising advances in recent years have led to theories (strings, 
branes, etc) which seem rather far from experimental tests. 

Although the quest for new physics finds its most convincing answer in the 
traditional  avenue of energies sufficient to produce and 
observe new particles, the lesson of recent decades has clearly 
indicated a second, indirect way of signalling the presence of new 
physics. We refer to the physics related to rare processes where Flavor 
Changing Neutral Currents (FCNC) and/or $CP$ violation occur.  As is 
well known, the first indications of the existence of charm and of a heavy 
top quark came before their direct discovery from evidence that indicated 
their role as virtual particles in $K$ and $B$ mixings, respectively. 
Obviously rare 
processes are a privileged ground for indirect signals of new physics. 
One can hardly believe that new physics contributions can be observable when 
considering phenomena that arise at the tree level in the SM. On
the contrary, FCNC and/or $CP$ violating processes offer the possibility for
loops containing new particles to compete with loops containing SM particles,
thus allowing  a fair chance for new
physics to emerge against the ``SM background''.

We think that the indirect road to new physics will become 
particularly relevant in the next few years. As we said, high energy 
accelerators have established the SM as the correct $100\ \mbox{GeV}$ theory, 
but they have not yet succeeded at the equally important task of observing 
some signals of new physics. After the end of LEP, our attention in  
high energy accelerator physics turns to the upgraded Tevatron at Fermilab. 
The evidence from the late activity at LEP in favor of  a light Higgs 
with a mass around $115\ \mbox{GeV}$  feeds our hopes 
that a Higgs signal may be detected at the Tevatron. Even if this were to
occur, we would have only an indirect indication that new physics must be far 
below some grand unification scale; the so--called ``big desert'' 
picture would be rejected (we remind the reader that if the SM is valid 
up to some very large scale close to the Planck scale the Higgs mass 
should exceed $130\ \mbox{GeV}$). The class of SM extensions that goes 
under the 
name of low energy supersymmetry (SUSY) would be favored, although they 
certainly do not exhaust the possibilities for new physics with a light 
Higgs. However, unless superpartners are light, it might be difficult 
to establish their existence at the Tevatron. If superpartners elude the
Tevatron, we will have to wait for the Large Hadron Collider (LHC) at CERN
to operate before deciding 
``directly'' whether low energy SUSY is a reality or just an intriguing 
intellectual construction.

What can we do to search for new physics in the meantime?.     
Here is where the FCNC and/or $CP$ violating processes come into full play 
in our challenging (sometimes desperate) effort to find hints of new 
physics. Two brand--new $B$ factories have just started operating, and there 
are plans for new experiments in rare kaon decays, in testing the electric 
dipole moments, in flavor violating leptonic processes, etc. The
enormous and interesting project of probing the SM in rare processes 
will proceed at full speed in the pre--LHC epoch. It is 
important to critically discuss strategies to look for new 
physics in such processes. Our goal here is to report on 
some of the most relevant aspects of the search for some  
signal of departure from the SM expectations in $CP$ violating 
phenomena. 
Admittedly, even if we find such signals it will be hard  
to establish what new physics is responsible for them. However, we  
indicate how, at least in the specific framework of low energy SUSY,
the combination of several observations may shed important light on 
the new physics, hence directing direct searches towards specific targets.
      
We refer to ``FCNC and/or $CP$ violation'' in qualifying the interesting 
processes for indirect detection of new physics because we can consider 
three different classes of them: (a) phenomena that require FCNC but occur 
even without $CP$ violation as $K$--$\bar{K}$ or $B$--$\bar{B}$ mixing; 
(b) $CP$ violating processes without FCNC, namely Electric Dipole Moments 
(EDMs) of the neutron, electron, etc; and finally,
(c) simultaneous FCNC and $CP$ violating phenomena, for instance the 
quantities $\varepsilon$ or $\varepsilon^\prime/\varepsilon$ in $K$ 
decays into pions and the $CP$ asymmetries in $B$ decays. 

In our view, among these three classes of rare processes, those 
related to the presence of $CP$ violation deserve a special attention 
for at least three reasons.

\begin{enumerate}
\item After nearly four decades of intensive experimental and theoretical 
work $CP$ violation still appears rather 
mysterious and, hence, a potentially good candidate to offer surprises in 
future tests (for general and recent reviews on $CP$ violation see 
\cite{booksbigiebranco}). We recently witnessed two major breakthroughs in our 
understanding of $CP$ violation. First, from the 
measurements \cite{epsilon'} of the 
$\varepsilon^\prime/\varepsilon$ parameters from both sides of the Atlantic,
we got the information that $CP$ violation occurs not only in $K$--$\bar{K}$
mixing ($\Delta S=2$), but also in the direct $K$ decay amplitudes 
($\Delta S=1$). 
Hence we can definitely reject the idea that $CP$ violation arises only from 
some specific superweak interactions in processes that change strangeness by 
two units. The second relevant piece of information, which we obtained last 
year, is that $CP$ violation is not present exclusively in the kaon system; 
there exists at least one other place where it shows up, namely in $B$ 
physics. This evidence emerged from the measurement of the $CP$ 
asymmetry in the decays of $B$ into $J/\psi K_s$ in three different 
experiments: $a_{J/\psi} = 0.34 \pm 0.20 \pm 0.05$ at BaBar \cite{babar},
$a_{J/\psi} = 0.58^{+0.32+0.09}_{-0.34-0.19}$ at BELLE \cite{belle} and
$a_{J/\psi} = 0.79^{+0.41}_{-0.44}$ at CDF \cite{CDF}. Interestingly enough, 
although the measurement coming from CDF confirms the SM expectation for such 
asymmetry, the results from the two abovementioned $B$ factories indicate 
central values significantly lower than the SM expectations. The 
errors are still quite large and so the best advice is the usual ``wait 
and see'', but this example illustrates how surprises in the 
$CP$ exploration can be lurking around the corner.

\item From the theoretical point of view, it is important to emphasize that 
new physics beyond the SM generically introduces new sources of $CP$ 
violation in addition to 
the usual CKM phase of the SM. Indeed, it is a common experience of model 
builders that if one tries to extend the SM with some  low energy new 
physics one must somehow control the proliferation of new 
$CP$ violating contributions. Significant portions of the parameter 
spaces of  new physics models can generally be ruled out by the severe 
constraints imposed by $CP$ violating phenomena
\cite{nirrattazzi}. For instance, as we discuss bellow, even if one 
considers the minimal supersymmetric extension of the SM that passes all the 
FCNC tests unscathed, one still faces severe problems in matching 
the experimental results concerning processes with $CP$ violation,
especially the constraints coming from the various bounds on electric 
dipole moments.

\item The third reason which makes us optimistic in having new physics 
playing a major role in $CP$ violation concerns the matter--antimatter 
asymmetry in the universe. Starting from a baryon--antibaryon symmetric 
universe, the SM is unable to account for the observed baryon asymmetry. 
The presence of new $CP$--violating contributions beyond the 
SM looks crucial to produce an efficient mechanism for the generation of a 
satisfactory $\Delta B$ asymmetry.
\end{enumerate}

The aim of this article is to discuss the above points with the goal of 
exploring the potentialities of $CP$ violation in our quest for new physics. 
Because of the vastness of the subject, we focus on a promising class 
of new physics that goes under the 
generic name of low energy SUSY \cite{reviewsusy}. We emphasize that low 
energy SUSY does 
not denote a well defined model; rather, it includes a variety of SM 
extensions (with a variety of phenomenological implications). We characterize 
this huge class of models according to their main features 
in relation to $CP$ violation. We want to avoid losing our 
readers in endless ``botanic'' classifications of SUSY models, but we
also want to avoid the idea that one can 
discuss $CP$ violation in ``the'' SUSY model as if one were discussing a 
specific construction such as the SM.

As appealing as low energy SUSY may appear, at least to some 
physicists, one should not forget that 
after more than twenty years of searches we do not have any experimental 
hint of SUSY particles. Furthermore, one could argue theoretically that the 
idea that superpartners appear on the TeV scale is motivated only by the 
hope that SUSY resolves the gauge hierarchy problem. The unification of the 
gauge couplings when low energy SUSY particles are included in their running 
can be considered only circumstantial evidence in favor of SUSY. 
In spite of all this, our choice to stick to low energy SUSY for most of 
this article in exploring the 
impact of $CP$ violation on new physics is justified by a simple and 
uncontroversial fact: we do not know of any other ``complete'' model of 
new physics that tackles the gauge hierarchy problem and successfully 
passes the impressive list of direct and indirect experimental tests of new 
physics. In any case, in the final part of this article we comment on $CP$ 
violation in other alternative possibilities for new physics.

Our discussion follows the general lines of distinction between
$CP$ violating processes with and without FCNC that we underlined above. In
the next Section we focus on $CP$ violation in the electric dipole moments
(hence without FCNC) and we introduce the first aspect of what is
called the ``SUSY $CP$ problem''. Then in Section 3 we move to flavor
changing $CP$ violation in SUSY and we deal with the second aspect of
the SUSY $CP$ problem. Section 4 is devoted to an overview of $CP$ violation
in other extensions of the SM. Finally in Section 5 we present our 
conclusions and outlook.
                             
\vskip .8cm
\section{SUSY $CP$ VIOLATION AND ELECTRIC DIPOLE MOMENTS}
Before entering the more specific discussion of the next sections, we offer 
a quick overview on $CP$ violation in SUSY at a more introductory level. 

When one decides to supersymmetrize the SM one has many options
\cite{reviewsusy}. First, 
the fields of the SM are embedded into superfields containing also the 
SUSY partners of each known particle; but how many superfields should one 
introduce? Here we stick to the minimal option: we introduce 
the minimal amount of superfields that are strictly demanded to obtain a 
viable supersymmetrization of the SM. This means that each particle will 
be accompanied by a superpartner, except in the Higgs sector 
where we have to introduce a second Higgs doublet in addition to the usual 
SM Higgs doublet. Some models introduce new singlet 
superfields or even more complicate structures, and such models 
may have new sources of $CP$ violation. A second important 
limitation in the 
class of SUSY models that we consider here concerns the imposition 
of $R$ parity. This discrete symmetry is usually added to the 
gauge and super--symmetries in order to prevent excessive baryon and 
lepton number violations. Although some symmetry must indeed be imposed 
to inhibit proton decay, $R$ parity is not the only way to 
achieve this. A vast class of alternative models imposes 
some discrete symmetry other than $R$, allowing for either 
baryon or lepton number violations. In such models, many new Yukawa 
couplings exist and then the number of $CP$ violating phases of the theory is 
sizeably increased. 

Even in minimal Supersymmetric versions of the SM (MSSM) 
where the minimal number of superfields is introduced and $R$ parity is 
imposed, one is still left with more than 100 free parameters, almost half 
of them given by $CP$ violating phases \cite{124}. 
Fortunately most of this huge parameter space is already 
phenomenologically ruled out. Indeed, FCNC and $CP$ violating processes play 
a major rule in drastically reducing the parameter space. Obviously it is 
difficult to make phenomenological predictions with so many 
free parameters, and so through the years many theoretical further 
restrictions have been envisaged for the MSSM class. The most drastic 
reduction on the SUSY parameter space leads to what is called the constrained 
MSSM (CMSSM) or minimal supergravity \cite{reviewsusy}. In the absence of 
phases, this model is characterized by only four parameters plus the 
sign of a fifth parameter.

In any MSSM, at least two new ``genuine'' SUSY $CP$--violating phases are
present. They originate from the SUSY parameters $\mu$, $M$, $A$ and $B$. 
The first of these
parameters is the dimensionful coefficient of the $H_u H_d$ term of the
superpotential. The remaining three parameters are present in the sector 
that softly breaks the $N=1$ global SUSY. $M$ denotes the common value of 
the gaugino masses, $A$ is the trilinear scalar coupling, and $B$ 
denotes the bilinear scalar coupling. In our notation, all these three 
parameters are dimensionful. The simplest way to see which combinations of 
the phases of these four parameters are physical \cite{Dugan} is to notice
that for vanishing
values of $\mu$,  $M$, $A$ and $B$ the theory possesses two additional    
symmetries \cite{Dimopoulos}. Indeed, if $B$ and $\mu$ are set to zero, a 
$U(1)$ Peccei--Quinn symmetry emerges, which in particular rotates 
$H_u$ and $H_d$.
If $M$, $A$ and $B$ are set to zero, the Lagrangian acquires a continuous
$U(1)$ $R$ symmetry. Then we can consider  $\mu$,  $M$, $A$ and $B$ as 
spurions that break the $U(1)_{PQ}$ and $U(1)_R$ symmetries. In this way, 
the question concerning the number and nature of the meaningful phases 
translates into the
problem of finding the independent combinations of the four parameters 
that are invariant under $U(1)_{PQ}$ and $U(1)_R$ and determining their 
independent phases. There are three such independent combinations, but 
only two of their phases are independent. We use here the commonly adopted 
choice:  
\begin{equation}
  \label{CMSSMphases}
  \phi_A = {\rm arg}\left( A^* M\right), \qquad
  \phi_\mu = {\rm arg}\left( \mu^* M\right).
\end{equation}
where also ${\rm arg}\left( B \mu\right) = 0$, i.e.
$\phi_B= - \phi_\mu$.

The main constraints on the SUSY phases come from their
contribution to
the electric dipole moments of the neutron and of the electron. For 
instance,
the effect of $\phi_A$ and $\phi_\mu$ on the electric and 
chromoelectric dipole moments of the light quarks ($u$, $d$, $s$) leads to a 
contribution to $d_n$ of order \cite{EDMN}

\begin{equation}
  \label{EDMNMSSM}
  d_n \sim 2 \left( \frac{100\ \mbox{GeV}}{\widetilde{m}}\right)^2 \sin 
\phi_{A,\mu}
  \times 10^{-23} {\rm e\, cm},
\end{equation}                               
where $\widetilde{m}$ here denotes a common mass for sleptons and gauginos. The
present experimental bound, $d_n < 1.1 \time 10^{-25}$ e cm, implies 
that $\phi_{A,\mu}$ should be $<10^{-2}$, unless one pushes SUSY masses up to
${\cal{O}}$(1 TeV). A possible caveat to such an argument calling for a
fine--tuning of $\phi_{A,\mu}$ is that uncertainties in the estimate of the 
hadronic matrix elements could relax the severe bound in 
Equation~\ref{EDMNMSSM} \cite{Ellis}.

These considerations lead most authors dealing with the MSSM
to simply put $\phi_A$ and $\phi_\mu$ equal to zero. Actually, one
may argue in favor of this choice by considering the soft breaking sector of 
the MSSM as resulting from SUSY breaking mechanisms that force $\phi_A$ 
and $\phi_\mu$ to vanish. For instance, it is conceivable that both $A$ and 
$M$ originate from one source of $U(1)_R$ breaking. Since $\phi_A$ 
measures the relative phase of $A$ and $M$, in this case it would 
``naturally'' vanish. In some specific models, it has been shown 
\cite{Dine} that through an analogous mechanism $\phi_\mu$ may also vanish. 

In recent years, the attitude towards the 
EDM problem in SUSY and the consequent suppression of the SUSY phases has 
significantly  changed. Indeed, options have been 
envisaged allowing for a conveniently suppressed SUSY contribution to the 
EDM even in the presence of large (sometimes maximal) SUSY phases. 
Methods of suppressing the EDMs consist of cancellation of various SUSY 
contributions among themselves \cite{cancel}, non universality of the soft 
breaking parameters at the unification scale \cite{non-u} and approximately 
degenerate heavy sfermions for the first two generations \cite{heavy}.
In the presence of one of these mechanisms, large supersymmetric phases 
are expected yet EDMs should be generally close to the experimental 
bounds. In the following we discuss the implications of these 
new $CP$ violating phases that may be present in SUSY, keeping in mind that 
some of the above mechanisms may be required to satisfy the constraints 
coming from the EDM's.

\section{FLAVOR CHANGING $CP$ VIOLATION}
\label{sec:FC$CP$V}
Despite of the large sensitivity of EDMs to the presence of new
phases, so far only neutral meson systems, $K^0$--$\bar{K}^0$ or
$B^0$--$\bar{B}^0$, show measurable effects of $CP$ violation. This fact
is, at first sight, surprising because in the neutral mesons $CP$ violation
is associated with a change in flavor and hence is CKM suppressed, whereas
EDMs are completely independent of flavor
mixing. The reason for this is that, in the SM, $CP$ violation is
intimately related to flavor, to the extent that observable $CP$
violation requires, not only a phase in the CKM mixing
matrix, but also three non--degenerate families of
quarks \cite{jarlskog}. As shown in the previous section, the
supersymmetrized SM contains new sources of $CP$, both flavor
independent or flavor dependent. Although the new phases are, in principle,
strongly constrained by the EDM experimental limits, we have seen
that several mechanisms allow us to satisfy these constrains with large
supersymmetric phases. Next, we analyze this possibility and the
effects in flavor changing $CP$ violation observables. We first 
concentrate on a MSSM with a flavor blind SUSY breaking, and
then we study a general MSSM in which the soft breaking terms include
all kinds of new flavor structures.

\subsection{Flavor Blind SUSY Breaking and $CP$ Violation}
\label{sec:flavor-blind}

The first step in our review of Supersymmetric $CP$ violation is the
analysis of a MSSM with flavor blind SUSY breaking. Flavor--blind
refers to a softly broken Supersymmetric SM in which the soft breaking
terms do not introduce any new flavor structure beyond the
Yukawa matrices whose presence in the superpotential is required to
reproduce correctly the fermion masses and mixing angles.
Supersymmetry is broken at a large scale, that we identify with
$M_{GUT}$, and from here, the parameters evolve with the standard MSSM
renormalization group equations (RGE) \cite{RGE,bertolini} down to the 
electro--weak scale. In these
conditions the most general allowed structure of the soft--breaking
terms at $M_{GUT}$ is,
\begin{eqnarray}
\label{soft}
& (m_Q^2)_{i j} = m_Q^2\  \delta_{i j} \ \ \  (m_U^2)_{i j} = m_U^2\  
\delta_{i j} \ \ \  
(m_D^2)_{i j} = m_D^2\  \delta_{i j} & \nonumber \\
&(m_L^2)_{i j} = m_L^2\  \delta_{i j} \ \ \  (m_E^2)_{i j} = m_E^2\  
\delta_{i j} \ \ \ \ \ 
m_{H_1}^2 \ \ \ \ \ \ m_{H_2}^2\ \ \ &\nonumber \\
& (A_U)_{i j}= A_U e^{i \phi_{A_U}} (Y_U)_{i j} \ \ \  
(A_D)_{i j}= A_D e^{i \phi_{A_D}}
(Y_D)_{i j} &\nonumber \\
& (A_E)_{i j}= A_E e^{i \phi_{A_E}} (Y_E)_{i j}. & 
\end{eqnarray}
where all the allowed phases are explicitly written except possible 
phases in the Yukawa matrices that
give rise to an observable phase in the the CKM matrix,
$\delta_{CKM}$. It is important to emphasize that, in this
flavor blind MSSM, $\delta_{CKM}$ is the only physical phase in the Yukawa
matrices and all other phases in $Y_U$ and $Y_D$ can be rephased away in 
the same way as in the SM. The absence of flavor structure in the scalar 
sector means that quarks and squarks can be rotated parallel already at the 
GUT scale and hence only $\delta_{CKM}$ survives. This is
not true in the presence of new flavor structures, where additional
Yukawa phases cannot be rephased away from quark--squark couplings 
\cite{piai}.
Furthermore, we also assume unification of gaugino masses at $M_{GUT}$
and the universal gaugino mass can always be taken as real.

The soft breaking terms structure in Equation~\ref{soft} includes, as the
simplest example, the CMSSM where all scalar
masses and $A$--terms are universal and the number of parameters is
reduced to 6 real parameters once we require radiative symmetry
breaking, ($m_{1/2}$, $m_0^2$, $A_0$, $\tan \beta$, $\phi_\mu$,
$\phi_A$) \cite{bertolini,CPcons,ibanezross}. More general soft--breaking 
terms in
the absence of new flavor structures can arise in GUT models
\cite{kawamura-murayama}. For instance, in a $SU(5)$ model, we expect
common masses for the particles in the ${\bf \bar{5}}$ and in the
${\bf 10}$ multiplets, and, in general, different masses for the two
Higgses. The new parameters in the soft--breaking sector would then be
($m_{1/2}$, $m_5^2$, $m_{10}^2$, $m_{H_1}^2$, $ m_{H_2}^2$, $A_{5}$,
$A_{10}$, $\tan \beta$, $ \phi_\mu$, $\phi_{A_5}$, $\phi_{A_{10}}$)
\cite{wien}. We take this structure as a representative example of
Equation~\ref{soft}, since it already shares all the relevant features.  In
any case, although the number of parameters is significantly increased
with respect to the CMSSM, it can still be managed and a full RGE
evolution and analysis of the low--energy spectrum is possible.

In this framework, we consider SUSY effects on flavor changing $CP$
violation and, in particular, the $CP$ asymmetry in the $b \to s \gamma$
decay, $\varepsilon_K$ and $B^0$ $CP$ asymmetries. However, we include
also two $CP$ conserving observables that are relevant in the fit of
the unitarity triangle, namely $\Delta M_{B_d}$ and $\Delta
M_{B_s}$. All these processes receive two qualitatively
different Supersymmetric contributions. As shown in the previous section,
Supersymmetry introduces new $CP$ violation phases that can
strongly modify these observables through their effects in SUSY
loops. On the other hand, even with vanishing SUSY phases, the
presence of the CKM phase in loops containing SUSY particles induces
new contributions that modify the SM predictions for these
observables.

Concerning the first possibility, we consider the following 
extreme situation: we analyze the effects of both $\phi_\mu$ and a
flavor independent $\phi_A$ in flavor changing $CP$ violation
experiments, ignoring completely (as a first step) EDM bounds. The result
looks rather surprising at first sight: 
{\it in the absence
of the CKM phase, a general MSSM with all possible phases in the
soft--breaking terms, but no new flavor structure beyond the usual
Yukawa matrices, can never give a sizeable contribution to
$\varepsilon_K$, $\varepsilon^\prime/\varepsilon$ or hadronic $B^0$ $CP$
asymmetries} \cite{flavor}. It is possible to understand the main 
reasons for this behavior (see \cite{flavor,CPcons} for a complete 
discussion). In a flavor blind
MSSM, all scalar masses are flavor universal at $M_{GUT}$. After RGE
evolution any off--diagonal entry in these matrices is necessarily
proportional to a product of at least two Yukawa matrix elements, which
finally can be translated into a quark mass squared and two CKM matrix 
elements, plus a loop factor. Thus, these
$LL$ or $RR$ sfermion matrices remain diagonal and real in very good
approximation.  A similar situation takes place in the trilinear
matrices: at $M_{W}$, they continue to remain proportional to the corresponding
Yukawa matrix up to small RGE corrections of the same kind, so they are nearly
diagonalized in the basis of diagonal quark Yukawas. In summary, the
situation in the sfermion mass matrices is exactly analogous to the
Constrained MSSM, and even though the $LR$ matrices are complex, flavor
off--diagonal elements are negligible.  Hence, gluino and neutralino
contributions are too small once we take into account the lower bounds
on squark and gaugino masses. The remaining possibility is chargino
loops where the necessary flavor change is provided by the usual CKM
mixing matrix.  In this case, we have both chirality conserving and
chirality changing transitions. It is easy to prove
\cite{fully,CPcons} that in the limit of flavor diagonal sfermion
masses and a real CKM matrix, the chirality conserving chargino
contributions are exactly real, irrespective of the phases in the
squark and chargino mixings. Therefore, only chirality changing
chargino loops could give rise to a Supersymmetric contribution
proportional to the new SUSY phases. Still, in the case of kaon
physics, these contributions are suppressed by the square of the $s$
quark Yukawa coupling and then completely negligible with respect to
the experimentally measured value. In the $B$ system we could argue
that the $b$ Yukawa coupling can be large \cite{fully}. However, these  
transitions are closely related to the decay $b \to s
\gamma$; constraints from this decay render the chargino chirality changing 
transitions negligible \cite{CPcons}. In other words, the effects of 
SUSY phases in a flavor blind MSSM are restricted
in practice to $LR$ transitions, such as the EDM or the $CP$ asymmetry 
in the $b\to s \gamma$ decay, and the effects in observables with dominant
chirality conserving contributions are negligible even with maximal
SUSY phases.

Accordingly, the most interesting $CP$ violation observable in these 
conditions is probably the $CP$ asymmetry in the $b \to s \gamma$ decay. 
However, we must take into account that the branching ratio itself
is a very strong constraint in any SUSY model \cite{bsg}. The $CP$ asymmetry 
is defined as,
\begin{eqnarray}
A_{CP}^{b\rightarrow s\gamma} &=&
\Frac{BR(\bar B\rightarrow X_s \gamma) - BR(B\rightarrow X_s \gamma)}
{BR(\bar B\rightarrow X_s \gamma) + BR(B\rightarrow X_s \gamma)}\nonumber\\
&\simeq&\Frac{1}{|C_7|^2} \left(0.012\, \mbox{Im}\{C_2 C_7^*\} - 0.093\, 
\mbox{Im}\{C_8 C_7^*\} \right. \nonumber \\
& &\left. +\  0.001\, \mbox{Im}\{C_2 C_8^*\} \right),
\end{eqnarray}
where the different $C_i$ are the Wilson coefficients of the current--current,
$Q_2=(\bar{s}_{L \alpha}\gamma_\mu c_{ L\alpha})\ (\bar{c}_{ L\beta} 
\gamma^\mu b_{ L\beta})$, magnetic, $Q_7=\Frac{e m_b}{16 \pi^2}\ 
\bar{s}_{L\alpha} \sigma_{\mu \nu} b_{R\alpha} F^{\mu \nu}$,
and chromomagnetic, $Q_8=\Frac{g_s m_b}{16 \pi^2}\
\bar{s}_{L\alpha} T_{\alpha \beta}^a \sigma_{\mu \nu} b_{R \beta} G^{a
\mu \nu}$, dipole operators, evaluated at the scale $\mu=m_b$
\cite{neubert}.  This asymmetry is predicted to be below $1\%$ in the
SM~\cite{soares,neubert}. On the other hand, we have seen that the
new SUSY phases can modify the $b \to s \gamma$ transition significantly. In
fact, several studies showed that the MSSM in the presence of large
$\phi_\mu$ and $\phi_A$ can enhance the $CP$ asymmetry up to $15 \%$
\cite{baek,aoki-cho,goto,hou,wien} which could be easily accessible at $B$
factories. In any case, it is important to remember that this scenario is 
viable only if some mechanism reduces the SUSY
contributions to the EDM. In the case of the CMSSM or a flavor blind MSSM
with possible EDM cancellations this analysis was repeated in
\cite{aoki-cho,goto,wien} and the asymmetry can reach at most a
few per cent. Figure 1 shows that without EDM 
constraints (open grey circles) the asymmetry can be above $5 \%$ at any value 
of the branching ratio and can reach even $13 \%$ for low branching ratios.
In Figure 1 the points of the 
parameter space that fulfill EDM constraints are represented by black dots.
The effect on the $CP$ asymmetry can be sizeable at low values of the 
branching ratio \cite{wien}, but for larger values of the branching ratio
the asymmetry is again around $1 \%$. In the plot, the points of the 
parameter space fulfilling EDM constraints are represented by black dots.
\begin{figure}
\begin{center}
\label{fig:acp}
\epsfxsize= 0.9 \linewidth
\epsfbox{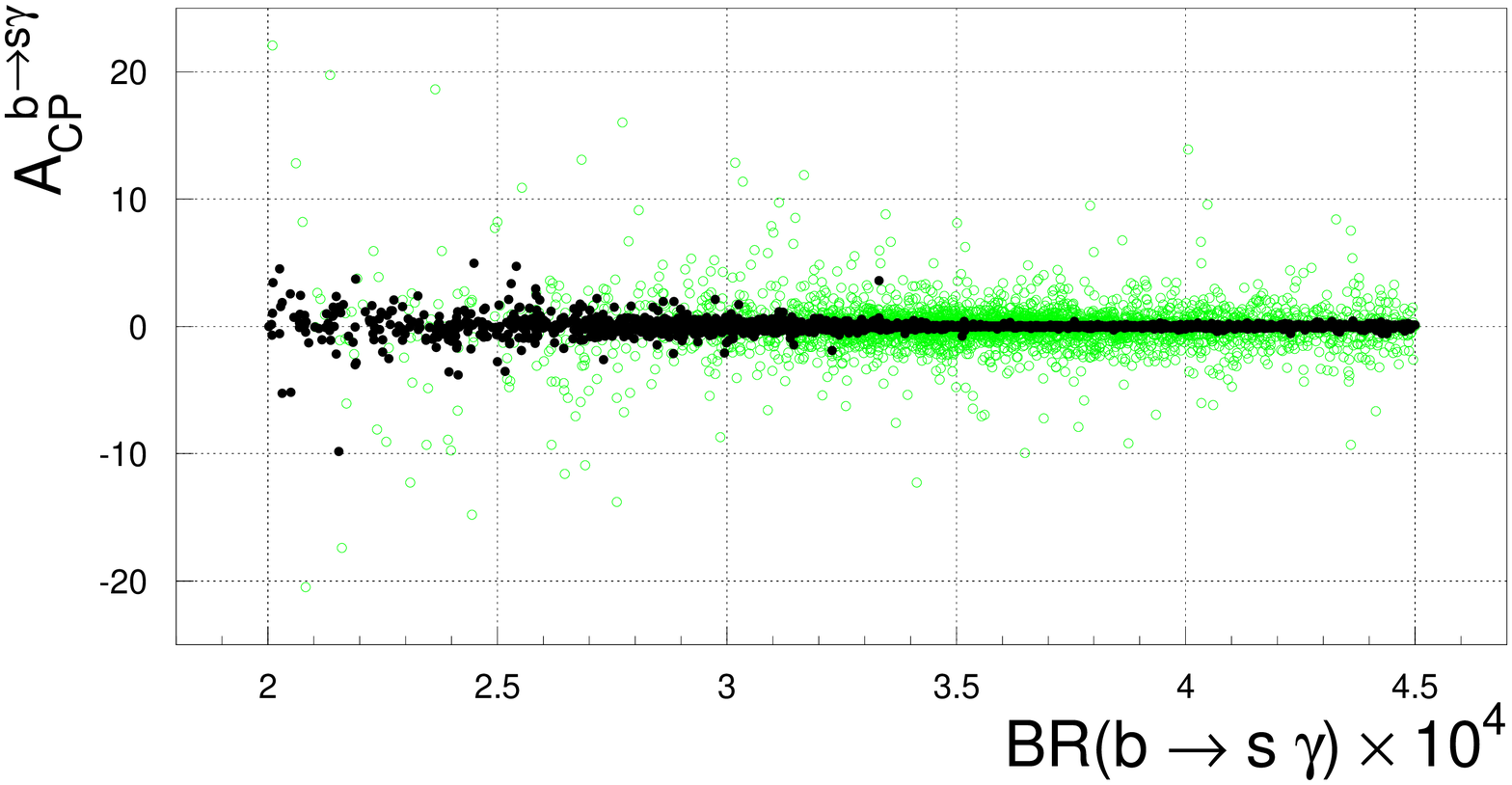}
\caption{$CP$ asymmetry in $\%$ versus branching ratio in the decay 
$b \to s\gamma$.
Black dots respect (or grey open circles violate) electric dipole moment 
constraints.}
\end{center}
\end{figure}

Still, even with $\phi_{A_i}=\phi_\mu=0$, SUSY loops can modify the
amplitudes proportional to the CKM phase. Indeed, in a flavor blind
MSSM, gluino and neutralino exchange contributions to FCNC are
subleading with respect to chargino ($\chi^\pm$) and charged Higgs
($H^\pm$) exchanges. This simply reflects the absence of flavor
mixing at tree level in gluino and neutralino contributions, whereas 
CKM mixing is present in the chargino and charged Higgs contributions.  
Hence, when dealing with $CP$ violating
FCNC processes in these models, we can confine the analysis to
$\chi^\pm$ and $H^\pm$ loops. This scenario was analyzed 
\cite{refbrignole,mpr,branco} soon after
the heavy top quark discovery opening the possibility of relatively
light stops, and more recently in Reference
\cite{MFV}. Specifically, these works analyze the effects of a Minimal
Flavor Violation  MSSM at the electroweak scale. That is, they
consider an MSSM with the CKM matrix as the only source of flavor
mixing even at the scale $M_W$. However, although they apply all the 
relevant low energy constraints, they do not consider the RGE effects
in the evolution of the soft--breaking terms and hence they take sparticle 
masses as completely unrelated.

Our point of view here is more restricted and we
follow the analysis of Reference \cite{wien}, where the flavor blind conditions
are specified at a large scale $M_{GUT}$ and the standard MSSM
RGEs \cite{2loop,RGE} are used to evolve the initial conditions down to
the electroweak scale. We consider two representative
examples of flavor blind MSSM, the CMSSM as the simplest model and the
$SU(5)$ inspired model defined above. Hence, we are mainly interested
in the following part of the low energy spectrum: $\chi^+$, $H^+$ and
$\tilde{t}$. Their masses are evolved to $M_W$ and then, all the relevant
experimental constraints are imposed:
\begin{itemize}
\item Absence of charge and color breaking minima and directions
unbounded from below \cite{casas}.
\item Lower bounds on masses from direct searches \cite{PDG}, in particular
$m_{\chi^+_i} > 90\ \mbox{GeV}$, $m_{\tilde{t}_i} > 90\ \mbox{GeV}$, 
$m_{\chi^0} > 33\ \mbox{GeV}$ and $m_{\tilde{\nu}} > 33\ \mbox{GeV}$.
\item Branching ratio of the $b \to s \gamma$ decay \cite{CLEO}.
\item Neutralino as the Lightest Supersymmetric Particle.
\end{itemize}
In this way, the complete supersymmetric spectrum at the electroweak
scale is obtained in terms of 6 or 11 parameters in the CMSSM or SU(5)
inspired model respectively. Within a MSSM scenario, this kind of analysis 
was first made in the work of Bertolini et al. \cite{bertolini} 
and has been updated several times since then \cite{others}. We follow
Bartl  et al. \cite{wien}, who developed a specialized study of the 
spectrum relevant for FCNC and $CP$ violation experiments. Indeed, the most
interesting point of this work is the strong correlation among
different SUSY masses that have a strong impact on low energy FCNC and
$CP$ violation studies.
\begin{figure}
\begin{center}
\label{fig:charstop}
\epsfxsize= 0.9 \linewidth
\epsfbox{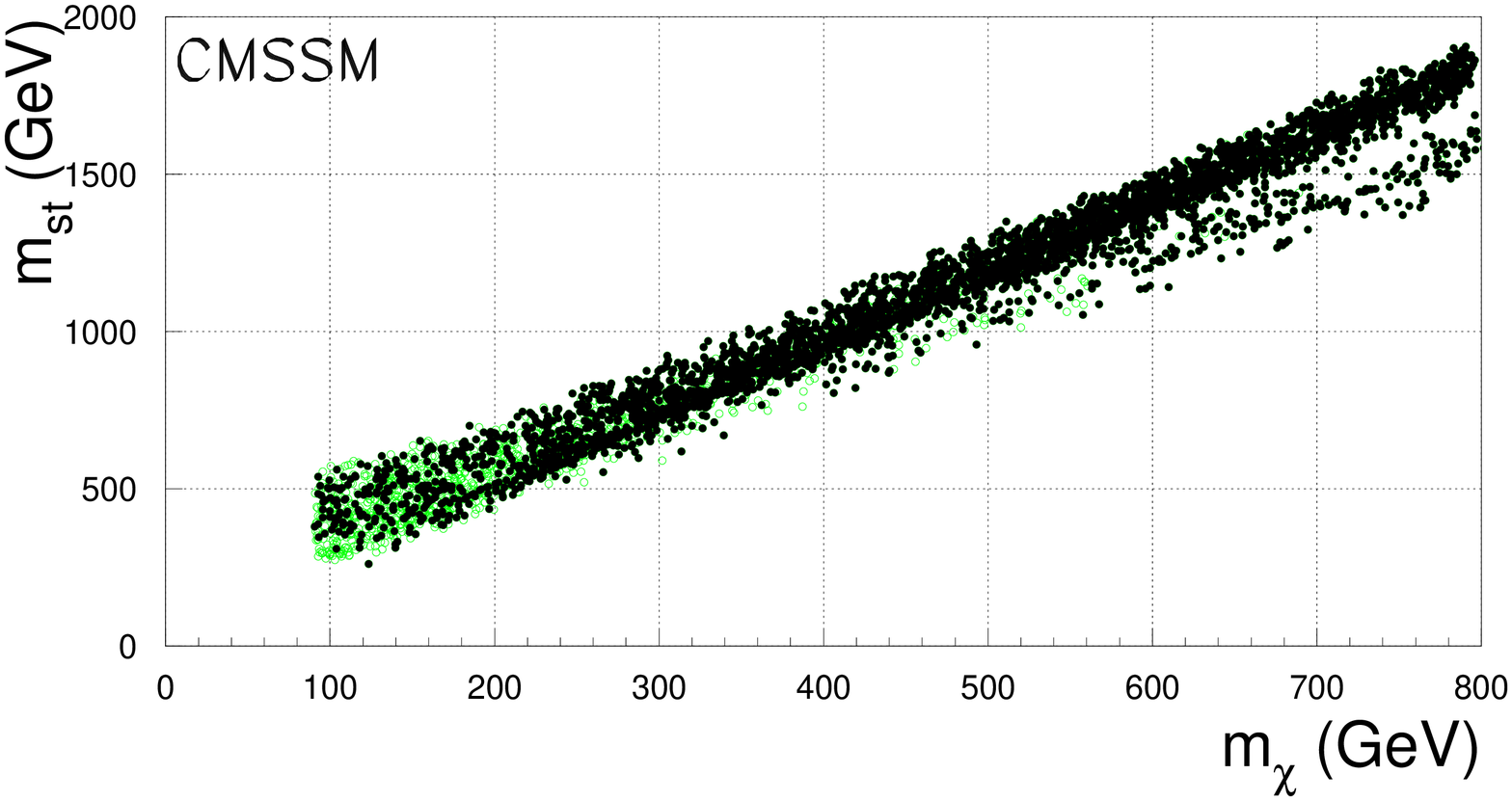}
\epsfxsize= 0.9 \linewidth
\epsfbox{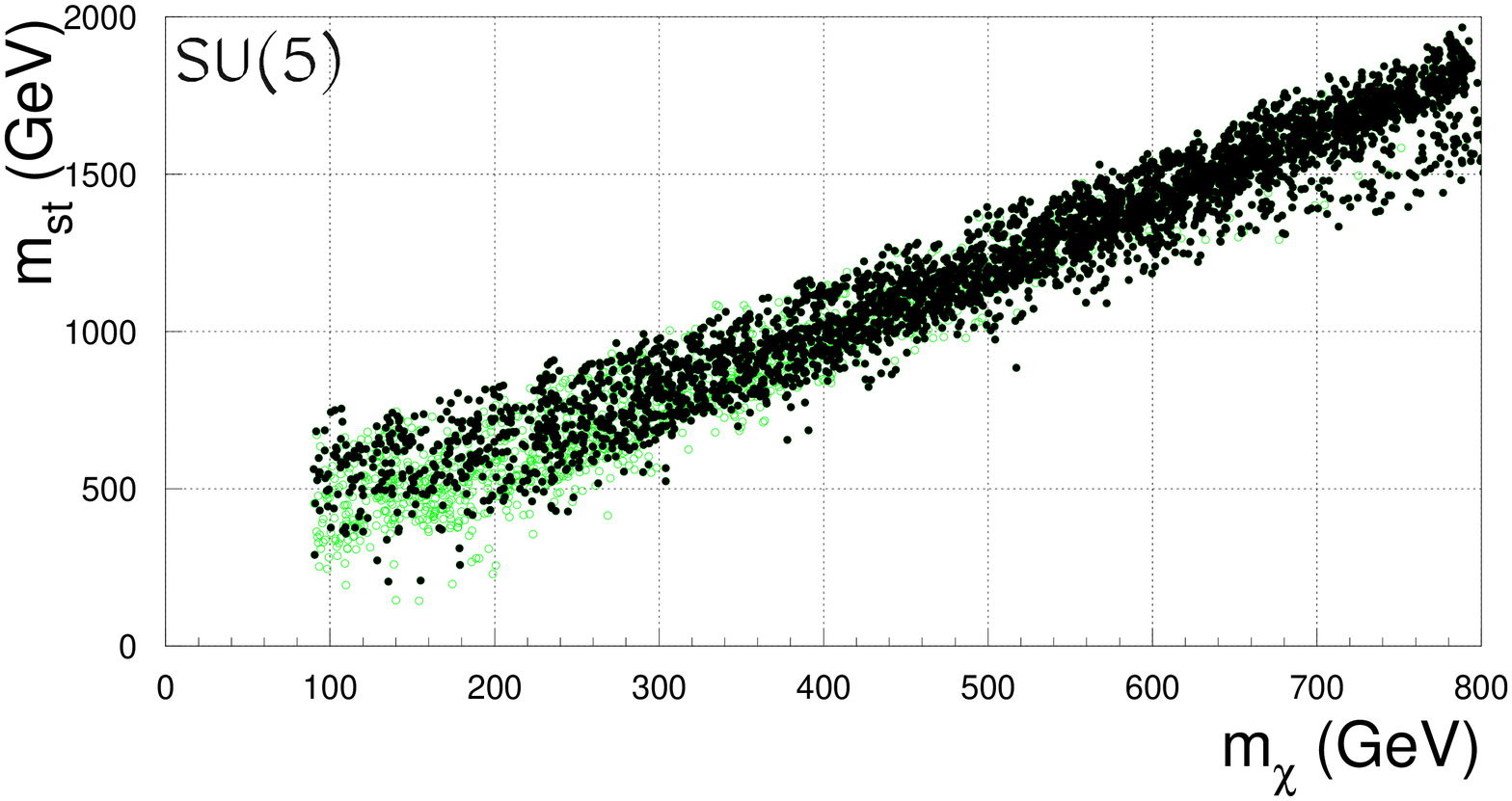}
\caption{Chargino mass versus lightest stop mass for the
parameter space described in the text in the CMSSM and SU(5) cases.
Black dots and open circles represent points satisfying or violating the 
$b \to s \gamma$ constraint respectively.}
\end{center}
\end{figure}
Figure 2 shows scatter plots of the mass of the
lightest chargino versus the lightest stop mass. In these plots we
vary the scalar and gaugino masses at $M_{GUT}$ as $100\ \mbox{GeV} < m_i <
1000\ \mbox{GeV}$, the trilinear terms as $0 < |A_i|^2 <
m_H^2+m_{\tilde{q}_L}^2+ m_{\tilde{q}_R}^2$ with arbitrary phases and 
$2 < \tan \beta < 50$.  It is interesting to notice
in this plot the very strong correlation among the chargino and stop
masses.  In fact, this correlation can be easily understood with the
help of the one--loop RGE \cite{RGE}. Neglecting for the moment the
so--called D-terms and the small radiatively generated
intergenerational squark mixing, we get for the stop masses:
\begin{eqnarray}
m^2_{{\tilde t}_{1,2}} =
    \frac{1}{2} \left( M^2_{Q_3} +  M^2_{U_3} + 2 m^2_t 
\mp \sqrt{ ( M^2_{Q_3} -  M^2_{U_3} )^2 + 4 m_t^2 (A_t - \mu \cot \beta)^2}
\right),
\label{eq:stopmass1}
\end{eqnarray}
in terms of the soft parameters at the electroweak scale.  Thanks to
the proximity of the top quark mass to its quasi--fixed point and the
relative smallness of $\mu \cot \beta$ for $\tan \beta \geq 2.5$ , we
can express Equation~\ref{eq:stopmass1} as a function of the initial
parameters at $M_{GUT}$ with only a small variation of the coefficients with
$\tan\beta$.  In the CMSSM case we find:
\begin{eqnarray}
m^2_{{\tilde t}_{1,2}} &=&
 0.43 M^2_0 + 4.55 M^2_{1/2} + m^2_t + 0.19\ \mbox{Re}(M^*_{1/2} A_0)
 \nonumber \\ &&
 \mp \Frac{1}{2} \sqrt{2.25 M^4_{1/2}  + 1.13 \, M^2_0 M^2_{1/2} + 20.2 \, 
m^2_t M^2_{1/2}}
\end{eqnarray}
Moreover, in the CMSSM $|\mu| \gsim \sqrt{3}\  m_{1/2}$ is always
larger than $M_2 \simeq 0.81\  m_{1/2}$ and hence, the lightest chargino
is predominantly gaugino. Then we can replace the initial gaugino mass
in terms of the lightest chargino mass and finally get,
\begin{eqnarray}
m^2_{{\tilde t}_{1,2}} &=&
 0.43 M^2_0 + 6.93 m^2_{\chi^+_1} + m^2_t + 0.23\ \mbox{Re}(m_{\chi^+_1} A_0)
 \nonumber \\ &&
\mp \Frac{1}{2}\sqrt{ 5.23 m^4_{\chi^+_1} + 1.72 \, M^2_0 m^2_{\chi^+_1}  + 
30.8 \, m^2_t m^2_{\chi^+_1}}
\end{eqnarray}
From this equation, we obtain for 
$100\ \mbox{GeV} < m_0 < 1\ \mbox{TeV}$ and with 
$m_{\chi}= 100\ \mbox{GeV}$ a maximal allowed range for the
lightest stop mass of $230\ \mbox{GeV} \lsim m_{{\tilde t}_1} \lsim 660\
\mbox{GeV}$.  As Figure 2 shows, this correlation is maintained for
larger chargino masses.  In the case of $SU(5)$, the main difference
is the fact that the Higgs masses are not tied to the other scalar
masses and now may be quite different. This has important effects on
the radiative symmetry breaking and in fact, lower values of $\mu$ are
possible such that the lightest chargino can have a predominant higgsino
component.  In the rare scenarios where $|\mu|\lsim M_2$, the stop masses
are somewhat lower than for the CMSSM case. Anyhow, as we can see
in Figure 2, a strong correlation is still maintained.  We must
emphasize that, due to gluino dominance in the soft--term evolution,
this kind of correlation is general in any RGE evolved MSSM from some
GUT initial conditions, assuming that gaugino masses unify as well.  In
summary, this implies that the ``light stop and chargino'' scenario in
any GUT evolved MSSM must be shifted to stop masses on the range of 
$250\ \mbox{GeV}$ and chargino masses of $100\ \mbox{GeV}$.  A very 
similar correlation can be obtained for other squark masses, roughly,
\begin{eqnarray}
m_{\tilde{q}} \simeq 9.3 \times m_\chi^2 + m_0^2 
\end{eqnarray}

Finally, we discuss the charged Higgs--boson mass. In Figure 3, 
we show a scatter plot of the mass of the charged Higgs
boson versus the value of $\tan \beta$ for the CMSSM and the SU(5)
examples. The main features here are the fact that, for low $\tan
\beta$, the masses of the charged Higgs--boson are above $400\ \mbox{GeV}$, and
that specially in the CMSSM case, most of these light Higgs--boson masses 
are eliminated by the $b \to s \gamma$ constraint.  
For larger values of $\tan \beta$
slightly lighter masses are allowed, but it is still true that we
seldom find charged Higgs masses below $300\ \mbox{GeV}$ in any case. The
reason for this is again the gluino dominance in the RGE. For
instance, at $\tan \beta=5$ the charged Higgs is $m_{H^+}^2\simeq 1.23
m_0^2 + 3.31 m_{1/2}^2$ and at $\tan \beta=30$, $m_{H^+}^2\simeq 0.72
m_0^2 + 1.98 m_{1/2}^2$, from the one--loop RGE \cite{wien}.
\begin{figure}[H]
\begin{center}
\label{fig:H+}
\epsfxsize= 0.9 \linewidth
\epsfbox{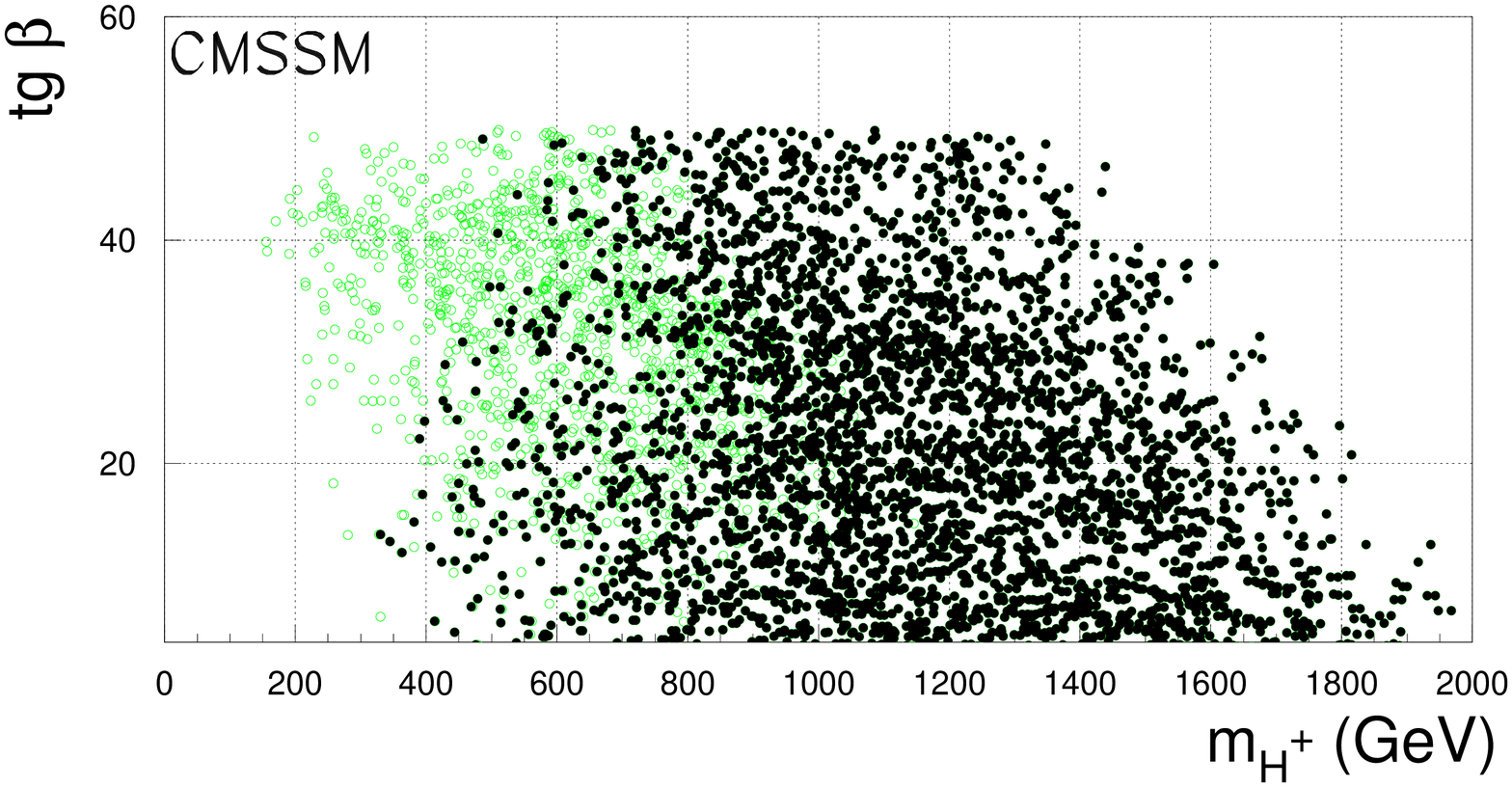}
\epsfxsize= 0.9 \linewidth
\epsfbox{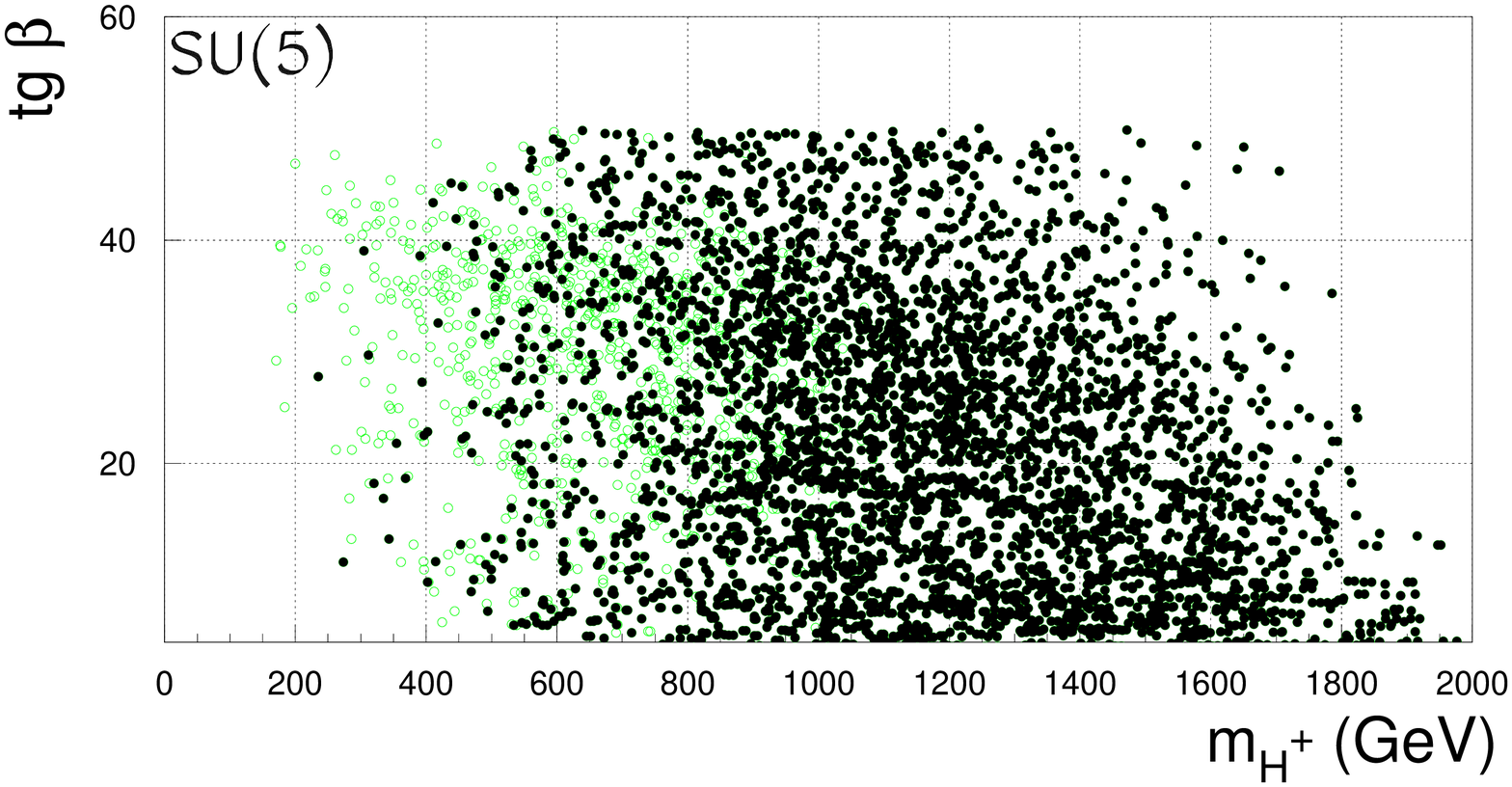}
\caption{Charged Higgs--boson mass as a function of $\tan \beta$ for the
parameter space described in the text in the CMSSM and SU(5) cases.
Black dots and open circles represent points satisfying or violating the 
$b \to s \gamma$ constraint respectively.}
\end{center}
\end{figure}
Taking into account the relevant features of the MSSM spectrum discussed above,
we can discuss the SUSY contribution in a flavor blind
scenario to the different $CP$ violating observables. First, the $b
\to s \gamma$ $CP$ asymmetry has already been discussed in the presence of 
large supersymmetric phases that survive the EDM constraints through a 
cancellation mechanism. In that case, the asymmetry could reach a few percent;
however, with vanishing SUSY phases we again obtain an asymmetry in the 
range of the SM value, well below $1\%$ \cite{neubert,soares}. A similar 
situation is found in $\varepsilon^\prime/\varepsilon$ where the SUSY
contributions tend to lower the SM prediction \cite{giudice,burasMFV}.

Second, the $\Delta F=2$ observables, i.e $\varepsilon_K$ and 
$B^0$--$\bar{B}^0$ mixing which play a fundamental role in the unitarity 
triangle fit, are also modified by new SUSY contributions.
Taking into account the new SUSY contributions, the SM fit of the unitarity 
triangle is modified and one obtains different restrictions on the $\rho$ 
and $\eta$ parameters of the CKM matrix. Moreover this fit has to be
compatible with the new direct measurements of the $B^0$ $CP$ asymmetries 
\cite{babar,belle}.  As explained elsewhere \cite{wien},
given that the SUSY contributions tend to interfere constructively
with the SM with a factorized CKM dependence, this implies that for a
given SUSY contribution the values of $\eta$ required to saturate
$\varepsilon_{K}$ are now smaller. The value of
$|V_{td} V_{tb}^*|$ required to saturate $\Delta M_{B_d}$ is
analogously decreased. Hence, it is evident that the addition of SUSY
tends to lower the values of $\eta$ and increase the values of $\rho$
in the fit, therefore reducing the actual value of $\beta$ in the
direction of the recent experimental results
\cite{babar,belle}. However, as shown by Buras \& Buras \cite{buras2}, 
in any minimal flavor violation
model at the electroweak scale, a strong correlation exists among the
$\Delta F =2$ contributions to $\varepsilon_K$ and $\Delta M_{B_d}$, 
and this allows only a small departure of $\sin 2 \beta$ from the SM
prediction. Still, different values of $\alpha$ and
$\gamma$ are, in principle, allowed. Nevertheless, as shown above, the relative
heaviness \cite{ellis,wien} of the SUSY spectrum implies that the
deviation from the standard model fit in these models tends to be small
for these angles.
In summary, a flavor blind MSSM can not generate large deviations from 
the SM expectations in the $B^0$ $CP$ asymmetries \cite{babar,belle}. 

\subsection{$CP$ Violation in the Presence of New Flavor Structures}
\label{sec:newflavor}

Flavor universality of the soft SUSY breaking is a strong assumption and 
is known not to be true in many supergravity and string--inspired models 
\cite{effectiv-non-u,soni-louis,string-SG,typeI}. In these models, a non 
trivial flavor 
structure in the squark mass matrices or trilinear terms is generically 
obtained at the supersymmetry breaking scale.
Hence, sizeable flavor off--diagonal entries appear in the squark
mass matrices, and new FCNC and $CP$ violation effects can be expected. 
In fact,  most of these flavor off--diagonal entries are severely 
constrained or even ruled--out by low energy FCNC and $CP$ violation 
observables. 

A very convenient parameterization of the SUSY effects in
these rare processes is the so--called mass insertion approximation 
\cite{MI}.
It is defined in the super CKM (SCKM) basis at the electroweak scale, 
where all the 
couplings of sfermions to neutral gauginos are flavor diagonal. In this 
basis, the sfermion mass matrices are not diagonal. The sfermion propagators, 
now flavor off--diagonal, can be expanded as a series 
in terms of $\delta = \Delta/\widetilde{m}^2$, where $\Delta$ denotes the
off--diagonal terms in the sfermion mass matrices, with $\widetilde{m}^2$ an 
average sfermion mass \cite{gabbiani}. As a result, FCNC and $CP$ violation 
constraints can be expressed as model--independent upper bounds on these 
mass insertions at the 
electroweak scale and they can readily be compared with the corresponding 
mass insertions calculated in a well defined SUSY model. 

A complete analysis of this kind was performed (see \cite{gabbiani,hagelin}
for a more complete discussion) and the constraints 
from $\Delta S=2$ processes were later updated \cite{MIupdate}. 
In the following, we present the phenomenological constraints from 
\cite{gabbiani,MIupdate}.

The main constraints on $CP$ violating mass insertions (with non--vanishing 
phases) come from 
$\varepsilon_K$, $\varepsilon^\prime/\varepsilon$ and EDMs, although 
``indirect'' constraints from $b \to s \gamma$ and $B$--$\bar{B}$ 
mixing are also relevant.
These constraints are presented in tables \ref{imds2} to \ref{redb2}.
It is important to emphasize the strong sensitivity of 
$\varepsilon^\prime/\varepsilon$ to $(\delta_{LR})_{12}$ and of
$\varepsilon_K$ to $(\delta_{LL})_{12}$, which implies that it is 
difficult to saturate both simultaneously with a single mass insertion 
\cite{varios,kanebrhlik}.

What message should we draw from the constraints in tables 
\ref{imds2}--\ref{redb2}?
First, it is apparent that FCNC and especially $CP$
violating processes represent a significant test for SUSY extensions of
the SM. Taking arbitrary sfermion mass matrices completely unrelated to
the fermion mass matrices would lead to mass insertions of order unity. 
Consequently, the first conclusion we draw from the small numbers in tables 
\ref{imds2}--\ref{redb2} is that there must be some close relation between 
the flavor structures of the sfermion and fermion sectors. 
Large portions of the parameter spaces of minimal SUSY models are completely 
ruled out thanks to the severity of the FCNC and $CP$ constraints.

But then an even more important question emerges after one stares at tables
\ref{imds2}--\ref{redb2}: Given the strong constraints from FCNC and $CP$ 
violating processes that we have already observed, can we still 
hope to see SUSY signals in other rare processes? 
In particular, restricting this question to $CP$ violation, can we 
still hope to find a significant disagreement with the SM expectations when we
measure $CP$ violation in various B decay channels?
Fortunately for us, the answer to this last question is yes.
For example, it has been shown \cite{ciuchini} that considering 
the $CP$ asymmetry in several B decay channels, which in the SM would give 
just the same answer (the angle $\beta$ of the unitarity triangle), it is 
possible to obtain different values when SUSY effects are switched on. 
SUSY contributions to some of the decay amplitudes can be as high as 70$\%$ 
with respect to the SM contribution, whereas other decay channels are not 
affected at all by the SUSY presence. Hence, assuming large $CP$ violating 
phases in SUSY, one could find discrepancies with the SM expectations that 
are larger than any reasonable theoretical hadronic uncertainty in the SM
computation. We refer the interested reader to Reference \cite{ciuchini} for 
a detailed discussion.
 
It is worth emphasizing that the above example shows that there is
still room for sizeable SUSY signals in $CP$ violating processes, but this
represents some kind of ``maximal hope'' of what we can expect from SUSY. In
other words, one takes the maximally allowed values of relevant $\delta$'s
to maximize the possible SUSY deviations from SM on $CP$ observables.
A different question is what we can ``typically'' expect in a SUSY
model. As we stressed in the introduction, no ``typical'' SUSY
model exists; what we call low--energy SUSY represents a vast
class of models. Yet it makes sense to try to identify some features of
minimal SUSY models where no drastic departures from flavor universality
are taken and to consider in this more restricted context what we can expect.
\begin{table}
 \begin{center}
 \begin{tabular}{||c|c|c|c||}  \hline \hline
  $x$ &
 ${\sqrt{\left|\mbox{Im}  \left(\delta^{d}_{L} \right)_{12}^{2}
\right|} }$ &
 ${\sqrt{\left|\mbox{Im}  \left(\delta^{d}_{LR} \right)_{12}^{2}
\right|} }$ &
 ${\sqrt{\left|\mbox{Im}  \left(\delta^{d}_{L} \right)_{12}\left(\delta^{d}_{R}
 \right)_{12}\right|} }$ \\
 \hline
 $
   0.3
 $ &
 $
2.9\times 10^{-3}
 $ & $
3.4\times 10^{-4}
 $ & $
1.1\times 10^{-4}
 $ \\
 $
   1.0
 $ &
 $
6.1\times 10^{-3}
 $ & $
3.7\times 10^{-4}
 $ & $
1.3\times 10^{-4}
 $ \\
 $
   4.0
 $ &
 $
1.4\times 10^{-2}
 $ & $
5.2\times 10^{-4}
 $ & $
1.8\times 10^{-4}
 $ \\ \hline \hline
 \end{tabular}
 \ \caption[]{Limits from $\varepsilon_K$ on 
$\mbox{Im}\left(\delta_{A}^{d}\right)_{12}\left(\delta_{B}^{d}\right)_{12}$, 
with $A,B=(L,R,LR)$ including next--to--leading--order QCD corrections and 
lattice $B$ parameters \cite{MIupdate}, for an average squark mass 
$m_{\tilde{q}}=500\ \mbox{GeV}$ and for different values of 
 $x=m_{\tilde{g}}^2/m_{\tilde{q}}^2$.}
 \label{imds2}
 \end{center}
 \end{table}

 \begin{table}
 \begin{center}
 \begin{tabular}{||c|c|c||}  \hline \hline
  $x$ & ${\left|\mbox{Im} \left(\delta^{d}_{L}  \right)_{12}
\right|} $ &
 ${\left|\mbox{Im} \left(\delta^{d}_{LR}\right)  _{12}\right| }$ 
\\ \hline
 $
   0.3
 $ &
 $
1.0\times 10^{-1}
 $ & $
1.1\times 10^{-5}
 $ \\
 $
   1.0
 $ &
 $
4.8\times 10^{-1}
 $ & $
2.0\times 10^{-5}
 $ \\
 $
   4.0
 $ &
 $
2.6\times 10^{-1}
 $ & $
6.3\times 10^{-5}
 $ \\ \hline \hline
 \end{tabular}
 \caption[]{Limits from 
$\varepsilon^{\prime}/\varepsilon < 2.7 \times 10^{-3}$ 
 on  $\mbox{Im}\left(\delta_{12}^{d}\right)$, for an average squark 
mass $m_{\tilde{q}}=500\ \mbox{GeV}$ and different values of 
 $x=m_{\tilde{g}}^2/m_{\tilde{q}}^2$.}
 \label{imds1}
 \end{center}
 \end{table}

 \begin{table}
 \begin{center}
 \begin{tabular}{||c|c|c|c||}  \hline \hline
 $x$ & $\left|\mbox{Im}\left(\delta^{d}_{LR} \right)_{11} \right|$ &
 $\left|\mbox{Im}  \left(\delta^{u}_{LR} \right)_{11} \right|$ &
 $\left|\mbox{Im}  \left(\delta^{l}_{LR} \right)_{11} \right|$\\
 \hline
 $
   0.3
 $ &
 $
2.4\times 10^{-6}
 $ & $
4.9\times 10^{-6}
 $ & $
3.0\times 10^{-7}
 $ \\
 $
   1.0
 $ &
 $
3.0\times 10^{-6}
 $ & $
5.9\times 10^{-6}
 $ & $
3.7\times 10^{-7}
 $ \\
 $
   4.0
 $ &
 $
5.6\times 10^{-6}
 $ & $
1.1\times 10^{-5}
 $ & $
7.0\times 10^{-7}
 $ \\ \hline \hline
 \end{tabular}
 \caption[]{Limits on $\mbox{Im} \left(\delta_{LR} \right)_{11}$ from electric 
dipole moments, for $m_{\tilde{q}}=500$ GeV and $m_{\tilde{l}}=100$ GeV.}
  \label{dipoles}
 \end{center}
 \end{table}

 \begin{table}
 \begin{center}
 \begin{tabular}{||c|c|c||}  \hline \hline
 $x$ & $\left|\left(\delta^{d}_{L} \right)_{23}\right| $ &
 $\left|  \left(\delta^{d}_{LR} \right)_{23}\right| $ \\
\hline
 $
   0.3
 $ &
 $
4.4
 $ & $
1.3\times 10^{-2}
 $ \\
 $
   1.0
 $ &
 $
8.2
 $ & $
1.6\times 10^{-2}
 $ \\
 $
   4.0
 $ &
 $
26
 $ & $
3.0\times 10^{-2}
 $ \\ \hline \hline
 \end{tabular}
 \caption[]{Limits on the $\left| \delta_{23}^{d}\right|$ from
 $b\rightarrow s \gamma$ decay for an average squark mass 
$m_{\tilde{q}}=500\ \mbox{GeV}$ 
and different values of $x=m_{\tilde{g}}^2/m_{\tilde{q}}^2$.}
 \label{tab:bsg}
 \end{center}
 \end{table}
\begin{table}
 \begin{center}
 \begin{tabular}{||c|c|c|c||}  \hline \hline
 $x$ & $\sqrt{\left|\mbox{Re}  \left(\delta^{d}_{13} \right)_{LL}^{2}\right|} $ 
 &
 $\sqrt{\left|\mbox{Re}  \left(\delta^{d}_{13} \right)_{LR}^{2}\right|} $ &
 $\sqrt{\left|\mbox{Re}  \left(\delta^{d}_{13} \right)_{LL}\left(\delta^{d}_{13}
 \right)_{RR}\right|} $ \\
 \hline
 $
   0.3
 $ &
 $
4.6\times 10^{-2}
 $ & $
5.6\times 10^{-2}
 $ & $
1.6\times 10^{-2}
 $ \\
 $
   1.0
 $ &
 $ 
9.8\times 10^{-2}
 $ & $
3.3\times 10^{-2}
 $ & $
1.8\times 10^{-2}
 $ \\
 $
   4.0
 $ &
 $
2.3\times 10^{-1}
 $ & $
3.6\times 10^{-2}
 $ & $
2.5\times 10^{-2}
 $ \\ \hline \hline
 \end{tabular}
 \caption[]{Limits on $\mbox{Re}\left(\delta_{A}\right)_{13}\left(
\delta_{B}\right)_{13}$, with $A,B=(L,R,LR)$ from $B^0$--$\overline{B}^0$ 
mixing, for an average squark mass
 $m_{\tilde{q}}=500\ \mbox{GeV}$ and for different values of 
 $x=m_{\tilde{g}}^2/m_{\tilde{q}}^2$.}
 \label{redb2}
 \end{center}
 \end{table}
In the following, we analyze a ``realistic'' non--universal MSSM, and compute
the ``reasonable'' expectations for the different mass insertions in 
this context. 
In first place, we define our generic MSSM through a set of four general
conditions:
\begin{enumerate}
\item {\bf Minimal particle content}: we consider the MSSM, with no 
additional particles from $M_W$ to $M_{GUT}$.
\item {\bf Arbitrary Soft--Breaking terms 
${\cal{O}}(m_{3/2})$}: The supersymmetry soft--breaking terms as given at 
the scale $M_{GUT}$ have a completely general flavor structure, but 
all of them are of the order of a single scale, $m_{3/2}$.
\item {\bf Trilinear couplings originate from Yukawa couplings}: Although
trilinear couplings are a completely new flavor structure they are related
to the Yukawas in the usual way: $Y^A_{i j} = A_{i j}\cdot Y_{i j}$, with
all $A_{i j}\simeq {\cal{O}}(m_{3/2})$.
\item {\bf Gauge coupling and gaugino unification at $M_{GUT}$} and RGE 
evolution of the different parameters from that scale.
\end{enumerate}

In this framework, any particular MSSM is completely defined, once we
specify the soft--breaking terms at $M_{GUT}$. We specify these soft--breaking
terms in the basis in which all the squark mass matrices, 
$M_{\widetilde{Q}}^2,M_{\widetilde{U}}^2, M_{\widetilde{D}}^2$, are diagonal. 
In this basis, the Yukawa matrices are, $ v_1\, Y_d = K^{D_L \, \dagger}\cdot 
M_d\cdot K^{D_R}$ and $ v_2\, Y_u = K^{D_L \, \dagger}\cdot K^\dagger\cdot 
M_u\cdot K^{U_R}$, with $M_d$ and $M_u$ diagonal quark mass matrices, $K$ the 
CKM mixing matrix and $K^{D_L}$, $K^{U_R}$, 
$K^{D_R}$ unknown, completely general, $3 \times 3$ unitary matrices. 

Although our analysis is completely general within this scenario \cite{KvsB}, 
we prefer to discuss a concrete example based on type--I \cite{typeI} string 
theory, (see \cite{EDMfree} for definition).\footnote{The case where
large flavor--independent soft phases may give a dominant contribution to 
$CP$ violation has been discussed elsewhere \cite{kanebrhlik}} 
In this particular example, gaugino masses, right--handed squarks, and 
trilinear terms are non--universal. Gaugino masses are, 
\begin{eqnarray}
\label{gaugino}
M_3 & = & M_1 = \sqrt 3 m_{3/2} \sin \theta  e^{-i\alpha_S}, \nonumber\\
M_2 & = &  \sqrt 3 m_{3/2} \cos \theta \Theta_1 e^{-i\alpha_1},
\end{eqnarray}
whereas the $A$--terms are obtained as 
\begin{eqnarray}
A_{1}&=& -\sqrt 3 m_{3/2}(\sin \theta e^{-i\alpha_S}+
\cos \theta \left[\Theta_1 e^{-i\alpha_1}- \Theta_3 e^{-i\alpha_3})\right]
\nonumber\\ 
A_{2}&=& -\sqrt 3 m_{3/2}\left[\sin \theta e^{-i\alpha_S}+
\cos \theta (\Theta_1 e^{-i\alpha_1}- \Theta_2 e^{-i\alpha_2})\right] 
\nonumber \\
A_{3}&=& -\sqrt 3 m_{3/2} \sin \theta e^{-i\alpha_S}=-M_3, 
\label{Aterms}
\end{eqnarray}
for the trilinear terms associated to the first, second and third generation
right--handed squarks respectively.
Here $m_{3/2}$ is the gravitino mass, $\alpha_S$ and $\alpha_i$ are 
the $CP$ phases of the $F$ terms of the dilaton field $S$ and 
the three moduli fields $T_i$, and $\theta$ and $\Theta_i$ are 
goldstino angles with the constraint, $\sum \Theta_i^2=1$.
Hence, the trilinear SUSY breaking matrix, $(Y^A)_{ij}=(Y)_{ij}(A)_{ij}$, 
itself is obtained as, 
\begin{equation}
Y^A = \left(\begin{array}{ccc}
 &  &  \\  & Y_{ij} &  \\  &  & \end{array}
\right) \cdot 
\left(\begin{array}{ccc}
A_{C^9_3} & 0 & 0 \\ 0 & A_{C^9_2} & 0 \\ 0 & 0 & A_{C^9_1} \end{array}
\right),
\end{equation}
in matrix notation \cite{restricted}.
\vskip 0.25cm

In addition, universal soft scalar masses for quark doublets and 
the Higgs fields are obtained, 
\begin{equation}
\label{doublets}
m_{L_i}^2=m_{3/2}^2 \left[1-{3 \over 2}\cos^2 \theta(1- 
\Theta_1^2)\right].
\end{equation}
And finally, the soft scalar masses for quark singlets are non--universal,
\begin{eqnarray}
\label{singlets}
m_{R_i}^2=m_{3/2}^2(1-3\cos^2 \theta\  T_i),
\end{eqnarray}
with $T_i=(\Theta_3^2,\Theta_2^2,\Theta_1^2)$.

To complete the definition of the model, we need to specify as well the 
Yukawa textures. The only available experimental information is the 
CKM mixing matrix and the quark masses. We choose our Yukawa texture 
following two simple assumptions : (a) the CKM mixing matrix originates from 
the down--type Yukawa couplings (as done in Reference \cite{murayama}) 
and (b) our Yukawa 
matrices are hermitian \cite{RRR}.
With these two assumptions we get $K^{D_L}=K$ and $K^{U_L}=1$.
However, it is important to emphasize that given that now $K^{D_L}$ and 
$K^{U_L}$ measure the
flavor misalignment between quarks and squarks, and that we already use
the rephasing invariance of the quarks to make $K_{CKM}$ real, we can 
expect new observable (unremovable) phases in the quark--squark 
mixings, and in particular in the first two generation sector. That is,
\begin{eqnarray}
K^{D_L}=\ \left(\begin{array}{ccc} 1 - \lambda^2/2 & \lambda \ e^{i \alpha}  
& A\ \rho^\prime\ \lambda^3 e^{i \beta} \\ 
- \lambda \ e^{-i \alpha} & 1 - \lambda^2/2 & A\ \lambda^2 \ e^{i \gamma}\\ 
A\ \lambda^3\ ( e^{- i (\alpha +\gamma)} - \rho^\prime\ e^{- i \beta})  & 
\ \ \ -A\ \lambda^2 \ e^{- i \gamma}  & 1 
\end{array}\right)
\label{Kdl}
\end{eqnarray}
to ${\cal{O}}(\lambda^4)$; $A$ and $\rho^\prime=|\rho +i \eta|$ are the usual 
parameters in the Wolfenstein parameterization, both ${\cal{O}}(1)$.
We must emphasize here that the observable phase in the CKM mixing matrix 
corresponds to the combination $\delta_{CKM}= \beta -\alpha-\gamma$; 
hence it is transparent that we can have a vanishing $\delta_{CKM}$ 
while 
being left with large observable phases in the SUSY sector \cite{piai}.
Hence, the Yukawa matrices are, $ v_1\, Y_d = K^{D_L \, \dagger}\cdot 
M_d\cdot K^{D_L}$ and $ v_2\, Y_u = M_u$. It is important to remember
that this is the simplest structure consistent with all phenomenological
constraints.

Now, the next step is to use the MSSM RGEs 
\cite{RGE,bertolini} to evolve these matrices down to the electroweak scale. 
The main RGE effects from $M_{GUT}$ to $M_W$ are those associated with the 
gluino mass and the large third generation Yukawa couplings. Regarding 
squark mass matrices, it is well--known that diagonal elements receive 
important RGE contributions proportional to gluino mass that dilute the 
mass eigenstate non--degeneracy, $m^2_{\widetilde{D}_{A_i}} (M_W) \simeq 
c_A^i \cdot m_{\tilde{g}}^2 + m_{\widetilde{D}_{A_i}}^2$ 
\cite{RGE,bertolini,CPcons}, with 
$c_L^{1,2}\simeq (6.5, 6.5)$, $c_L^3 \simeq (5.5, 4.6)$, 
$c_R^{1,2}\simeq (6.1,6.1)$ and $c_R^3 \simeq (6.1, 4.3)$ for 
$(\tan \beta=2.5, \tan \beta=40)$ \cite{wien}. 
In the SCKM basis, the off--diagonal elements in the sfermion mass matrices
are given by 
$(K^{A} \cdot M_{\widetilde{D}_A}^2 \cdot K^{A\,\dagger})_{i\neq j}$ up to 
smaller RGE corrections.
Similarly, gaugino effects in the trilinear 
RGE are always proportional to the Yukawa matrices, not to the trilinear 
matrices themselves and so they are always diagonal to extremely good 
approximation in the SCKM basis. Once more, the off--diagonal elements will 
be approximately given by 
$(K^{Q_L} \cdot Y^{A}_{Q} \cdot K^{Q_R \,\dagger})_{i\neq j}$. 

The $LR$ and $RR$ mass insertions are defined as $(\delta_{LR}^Q)_{i j}=
(M_{\widetilde{Q}_{LR}}^2)_{i j} /m^2_{\tilde{q}}$ and $(\delta_{RR}^Q)_{i j}=
(M_{\widetilde{Q}_{RR}}^2)_{i j} /m^2_{\tilde{q}}$ respectively.
Hence, in our example defined in 
Equations~\ref{gaugino}--\ref{Kdl}, we have $LR$ and $RR$ 
off--diagonal mass insertions, which can be estimated as,
\begin{eqnarray}
\label{Dlr}
(\delta_{LR}^{d})_{i j}= \frac{1}{m^2_{\tilde{q}}}\ m_i\ \Big[
K^{D_L}_{i 2}\ K^{D_L\, *}_{j 2}\ ( A_{2}^* - A_{1}^* ) +
K^{D_L}_{i 3}\ K^{D_L\, *}_{j 3}\ ( A_{3}^* - A_{1}^* ) \Big]
\end{eqnarray}
and
\begin{eqnarray}
\label{Dr}
(\delta^{d}_{R})_{i j}&\ =\ \Frac{1}{m^{2}_{\tilde{q}}}\ \Big[
 K^{D_L}_{i 2} K^{D_L\, *}_{j 2}\ (m_{R_{2}}^2 - m_{R_{1}}^2 )\
 +\  K^{D_L}_{i 3} K^{D_L\, *}_{j 3}\ (m_{R_{3}}^2 - 
m_{R_{1}}^2 )\ \Big]
\end{eqnarray}

Equation~\ref{Dlr} reveals an important feature of the $LR$ mass insertions.
Because of the trilinear terms structure in generic models of soft--breaking, 
the $LR$ sfermion matrices are always suppressed by $m_{q_i}/m_{\tilde{q}}$, 
with $m_{q_i}$ the mass of one of the quarks involved in the coupling 
and $m_{\tilde{q}}$ the average squark mass
\cite{restricted}. In any case, this suppression is necessary to avoid 
charge and color breaking and directions unbounded from below \cite{casas}. 
We can easily estimate the different mass insertions with these formulas. 
First we must take into account that, owing to the gluino dominance in the 
squark eigenstates at $M_W$, $ m^{2}_{\tilde{q}}(M_W)\approx 6\ 
m_{\tilde{g}}^2(M_{GUT})$. In the kaon system, 
we can neglect $m_d$; replacing the values of masses and mixings in
Equations~\ref{gaugino}--\ref{Kdl} we obtain, 
\begin{eqnarray}
\label{LRKaon}
(\delta^{d}_{LR})_{1 2}&\simeq& \Frac{m_s}{m_{\tilde{q}}}\ 
\Frac{(A_2 -  A_1)}{m_{\tilde{q}}}\ K^{D_L}_{1 2} K^{D_L\, *}_{2 2} \nonumber 
\\ 
&\simeq& 2.8 \times 10^{-5} \cdot (\Theta_2 e^{-i \alpha_2} - 
\Theta_3 e^{-i \alpha_3})\cdot \left(\Frac{100\ \mbox{GeV}}{m_{3/2}}  \right)
\end{eqnarray} 
where we have used $\theta \simeq 0.7$ as in Reference \cite{EDMfree}. 
Comparing this value with the bounds in Table~\ref{imds1}, we see 
that it could indeed give a very sizeable contribution to 
$\varepsilon^\prime/\varepsilon$ \cite{murayama,EDMfree,newflavor}. The phases 
$\alpha_2$ and $\alpha_1$ are actually unconstrained by 
EDM experiments as emphasized in \cite{EDMfree}.
This important result means that even if the relative 
quark--squark flavor misalignment is absent and the only flavor mixing is
provided by the usual CKM matrix, i.e. $K^{D_L}=K_{CKM}$, the presence of 
non--universal flavor--diagonal trilinear terms is enough to generate large 
FCNC effects in the kaon system.  

Similarly, in the neutral $B$ system, $(\delta^{d}_{LR})_{1 3}$ contributes 
to the $B_d-\bar{B}_d$ mixing parameter, $\Delta M_{B_d}$.
However, in our minimal scenario, $K^{D_L}\approx K$, we obtain,
\begin{eqnarray}
\label{LRbCKM}
(\delta^{d}_{LR})_{1 3} &\simeq& \Frac{m_b}{m_{\tilde{q}}}\ 
\Frac{(A_3 -  A_1)}{m_{\tilde{q}}}\ K^{D_L}_{1 3} K^{D_L\, *}_{3 3} \nonumber\\
&\simeq&
 2.5 \times 10^{-5} \cdot (\Theta_1 e^{-i \alpha_1} - 
\Theta_3 e^{-i \alpha_3})\cdot \left(\Frac{100\ \mbox{GeV}}{m_{3/2}} \right),
\end{eqnarray}
clearly too small to generate sizeable $\tilde{b}$--$\tilde{d}$
transitions, as the bounds in Table~\ref{redb2} show. Notice that 
larger effects are still possible in a
more ``exotic'' scenario with a large mixing in $K^{D_L}_{1 3}$.
For instance, with a maximal value, $|K^{D_L}_{1 3} K^{D_L\, *}_{3 3}| = 1/2$, 
we would get
$(\delta^{d}_{LR})_{1 3} \simeq 2 \times 10^{-3} \cdot (100\ \mbox{GeV} / 
m_{3/2})$.
Even in this limiting situation, this result is roughly one order of
magnitude too small to saturate $\Delta M_{B_d}$, though it could still be 
observed through the $CP$ asymmetries. Hence in the $B$ system we reach a 
very different result: it is not enough to have non--universal 
trilinear terms, large flavor misalignment among quarks and squarks
is also required.   

A similar analysis can be maid with the chirality conserving mass insertions.
From Equation~\ref{Dr}, in the kaon system, we get, 
\begin{eqnarray}
\label{dR12}
(\delta^{d}_{R})_{1 2}&\simeq& \Frac{\cos^2 \theta (\Theta_1^2 - 
\Theta_2^2)}{ 6 \sin^2 \theta}K^{D_L}_{1 2} {K^{D_L\, *}_{2 2}}\ +
\Frac{\cos^2 \theta (\Theta_1^2 - \Theta_3^2)}{ 6 \sin^2 \theta}K^{D_L}_{1 3} 
{K^{D_L\, *}_{2 3}} \nonumber \\ 
&\simeq & \Frac{\cos^2 \theta (\Theta_1^2 - \Theta_2^2)}{ 6 \sin^2 \theta} \ 
\lambda\ e^{i \alpha}
\end{eqnarray}
This value has to be compared with the mass insertion bounds required to 
saturate 
$\varepsilon_K$ \cite{MI}, which in this case are, 
$(\delta^{d}_{R})_{1 2}^{bound} \leq 0.0032$. Using $\theta \simeq 0.7$, 
we get,
\begin{eqnarray}
\label{result}
(\delta^{d}_{R})_{1 2}&\simeq& 0.035 (\Theta_1^2 -\Theta_2^2) \sin \alpha
\lsim 0.0032.
\end{eqnarray}
Hence, it is clear that we can easily saturate $\varepsilon_K$ without 
any special fine--tuning. Indeed, this constraint which is one of the main 
sources of the so--called Supersymmetric flavor problem, in this generic
MSSM amounts to the requirement that $(\Theta_1^2 -\Theta_2^2) \sin \alpha 
\lsim 0.1$ with all the different factors in this expression $\Theta_1^2,
\Theta_2^2,\sin \alpha \leq 1$ \cite{piai}.

Now we turn to the $CP$ asymmetries in the $B$ system. Once more, with 
Equation~\ref{Dlr} we have,
\begin{eqnarray}
\label{dR13}
(\delta^{d}_{R})_{1 3}&\simeq& \Frac{\cos^2 \theta (\Theta_2^2 - 
\Theta_1^2)}{ 6 \sin^2 \theta}K^{D_L}_{1 2} {K^{D_L\, *}_{3 2}}\ +
\Frac{\cos^2 \theta (\Theta_3^2 - \Theta_1^2)}{ 6 \sin^2 \theta}K^{D_L}_{1 3} 
{K^{D_L\, *}_{3 3}} \nonumber\\ 
&\simeq&  A\ \lambda^3 \ \Frac{\cos^2 \theta}{6 \sin^2 \theta}\left[ - 
(\Theta_2^2 - \Theta_1^2)\ e^{i (\alpha + \gamma)} \right. \nonumber \\
&& \left.+ (\Theta_3^2 - \Theta_1^2)
\ ( e^{- i (\alpha +\gamma)} - 
\rho\ e^{- i \beta})\right] \lsim 10^{-3},
\end{eqnarray}
to be compared with the mass insertion bound 
$(\delta^{d}_{R})_{1 2}^{bound} \leq 0.098$ required to not over--saturate the 
$B^0$ mass difference. 

We conclude that large effects are expected in the kaon system 
in the presence of non--universal squark masses even with a ``natural'' 
CKM--like mixing both for chirality changing and chirality conserving
transitions. The $B$ system is much less sensitive 
to supersymmetric contributions, so observable effects are expected only with 
approximately maximal $\tilde{b}$--$\tilde{d}$ mixings.

Recently, the arrival of the first measurements of $B^0$ $CP$ asymmetries 
from the $B$ factories has caused a great excitement in the high energy 
physics community.
\begin{equation}
\label{Bfactories}
\begin{array}{ll}
& \\
a_{J/\psi} & = \\ 
& \\
\end{array}
\left\{ 
\begin{array}{ll}
0.34\pm 0.20\pm 0.05 & (\mbox{Babar \cite{babar}}) 
\\0.58 ^{+0.32+0.09}_{-0.34-0.10}
& \left( \mbox{Belle \cite{belle}}\right) \\ 
0.79^{+0.41}_{-0.44}
& \left( \mbox{CDF \cite{CDF}}\right)
\end{array}
\right.  \label{3ajk}
\end{equation} 
The errors are still too large to draw any firm conclusion. Still, these
measurements, and specially the BaBar value which is the most precise one, 
leave room for an asymmetry considerably smaller than the standard 
model expectations corresponding to 
$0.59\leq a_{J/\psi}^{SM}=\sin \left( 2\beta \right) \leq 0.82$. 
This possible discrepancy, if confirmed, would be a first sign of the presence
of new physics in $CP$ violation experiments. 
Several papers have discussed the possible implications of a small $CP$ 
asymmetry
\cite{BBtheory,KvsB} and pointed out two possibilities. A small asymmetry
can be due to a large new physics contribution in the $B$ system and/or
to a new contribution in the $K$ system modifying the usual determination of 
the unitarity triangle. Taking into account the results above, in a
non--universal MSSM it is realistic to reproduce the $CP$ violation in the 
kaon system through SUSY effects, while being left with a small $a_{J/\psi}$ 
in the $B$ system. Indeed the role of the CKM phase could be confined to the SM
fit of the charmless semileptonic $B$ decays and
$B^0_d$--$\bar{B^0_d}$ mixing, while predominantly attributing to SUSY
the $K$ $CP$ violation ($\varepsilon_K$ and
$\varepsilon^{\prime}/\varepsilon$). In this case the CKM phase can be
quite small, leading to a lower $a_{J/\psi}$ $CP$ asymmetry \cite{piai}.

\section{$CP$ VIOLATION AND OTHER SM EXTENSIONS}

So far, we have reviewed $CP$ violation in SUSY as a particularly appealing 
extension of the SM. In this last section we briefly outline other SM 
extensions ( see \cite{nirrattazzi,nirnew} for a more complete 
discussion).

We classify other extensions of the SM with respect to its low energy spectrum.
In this way, we distinguish between models with an extended fermion sector, 
an extended scalar sector and an extended gauge sector. 
  
The fermion sector can be extended either by a complete fourth generation
or simply by the addition of extra nonsequential quarks \cite{vlq}.
A chiral fourth generation model faces strong restrictions from neutrino
masses and electroweak precision data \cite{PDG}. On the other hand, 
extensions with additional down quarks in a vector--like representation
of the SM are free from these problems and are especially interesting from
the $CP$ phenomenology point of view. These models naturally arise, for 
instance, as the low energy limit of an $E_{6}$ grand unified theory. 
At a more phenomenological level, models with isosinglet quarks provide the 
simplest self--consistent framework to study deviations of $3\times 3$ 
unitarity of the CKM matrix, as well as FCNC at tree level.
The extra vector--like quarks mix with the three ordinary down quarks giving 
rise to a $4\times 4$ mixing matrix, out of which the $3\times 4$ upper 
submatrix corresponds to the CKM matrix in charged currents. In this extended 
matrix, new observable phases arise and, through unitarity, produce 
complex tree level FCNC couplings in the $Z$ and Higgs vertices.
The fact that the experimental bounds on CKM mixings allow a larger mixing
with the third generation quarks implies that despite the tight 
constraints from other FCNC processes \cite{boundsVL}, these tree level 
FCNCs are able to modify significantly the SM prediction for $B^0$ $CP$ 
asymmetries \cite{B-VL}. Indeed, these models are even able to reproduce 
low asymmetries in the lower range of the recent BaBar result \cite{TLFCNC}.   

Many SM extensions, include additional scalar multiplets \cite{hunter}.
These models have additional Yukawa couplings, as well as scalar 
self--interactions that introduce many new sources of $CP$ violation. 
However, a generic model suffers from severe phenomenological problems 
that are solved with the requirement of extra symmetries \cite{NFC,AFS,HS}.
For example, the Higgs sector of the MSSM, discussed in previous 
sections is an example of a two--Higgs doublet model with natural flavor 
conservation \cite{NFC}, this means that each Higgs doublet couples to a 
different fermion sector. In this case, the only new ingredient is the 
presence of a physical charged Higgs with CKM couplings. In some cases, 
it is also possible that the neutral scalars are mixtures of $CP$ even and $CP$
odd fields \cite{CPoddmix}. This mixing shows up mainly in
EDM's, $b \to s \gamma$, and top quark decays.

An enlargement of the gauge group to include, for instance a left--right
symmetric theory is also a very attractive extension of the SM \cite{LRsym}.
Concerning $CP$ violation, it is especially interesting that these 
models can accommodate spontaneous $CP$ violation \cite{LRspont}. 
The relative phases among neutral Higgs vacuum expectation 
values are the only source of $CP$ violation here.
Actually, a single phase enters the charged fermion mass matrices giving rise
to six physical phases in these matrices. This is the only source 
for $CP$ violation at low energy, in a very predictive scheme. In fact, 
the rephasing invariant phase of the (left--handed) CKM matrix is small 
\cite{GBkaon}.
Hence, an extra contribution from right--handed bosons is required to saturate 
the experimental bound in $\varepsilon_K$. This, in turn, translates into 
an upper bound on the
mass of the right--handed boson. In a similar way, there can be sizeable new 
contributions in $B$--$\bar{B}$ mixing from the phases in the right--handed
mixing matrix and consequently, the $CP$ asymmetries in $B$ decays may be 
significantly different from the SM expectations \cite{GBbottom}.

\section{CONCLUSIONS AND OUTLOOK}

Here we summarize the main points in our review:
\begin{itemize}
\item There exist strong theoretical and ``observational'' reasons to go
beyond the SM.
\item The gauge hierarchy and coupling unification problems favor the presence 
of low--energy SUSY (either in its minimal version, CMSSM, or more naturally,
in some less constrained realization).
\item Flavor and $CP$ problems constrain low--energy SUSY, but, at the same
time, provide new tools to search for SUSY indirectly.
\item In general, we expect new $CP$ violating phases in the SUSY
sector. However, these new phases are not going to produce sizeable
effects as long as the SUSY model we consider does not exhibit a new flavor 
structure in addition to the SM Yukawa matrices.
\item In the presence of a new flavor structure in SUSY, large 
contributions to $CP$ violating observables are indeed possible.
\item $CP$ violation is also very sensitive to the presence of other low 
energy extensions of the SM.
\end{itemize}

Maybe a reader who has followed us up to this point would like to have a final 
assessment about the chances of having some signals of the presence of low 
energy SUSY in FCNC and $CP$ violating processes. It should be clear from our 
discussion that the answer heavily depends on the presence or absence of a 
link between the mechanism which is responsible for the flavor structure of the
theory and the mechanism originating the breaking and transmission of SUSY.

If no link is present (flavor blind SUSY), there exists  quite a few especial 
places where we can hope to ``see'' SUSY in action: the EDMs, the 
$A_{CP}^{b \to s \gamma}$, and, as it emerged recently, the anomalous
magnetic moment of the muon. On the other hand, in the more general (and,
in our view, also more likely) case where, indeed, SUSY breaking is not 
insensitive to the flavor mechanism, there exist a rich variety of FCNC
and $CP$ violation potentialities for SUSY to show up. As we have seen, 
$K$ and $B$ physics offer appealing possibilities: $\varepsilon_K$,
$\varepsilon^\prime/\varepsilon$, $CP$ violating rare kaon decays,
$CP$ asymmetries in $B$ decays, rare $B$ decays\dots . In fact, we think that
the relevance of SUSY searches in rare processes is not confined to the usually
quoted possibility that indirect searches can arrive ``first'', before direct 
searches (Tevatron and LHC), in signaling the presence of SUSY.
Even after the possible direct production and observation of SUSY particles,
the importance of FCNC and $CP$ violation in testing SUSY remains of utmost
relevance. They are and will be complementary to the Tevatron and LHC 
establishing low energy supersymmetry as the response to the electroweak 
breaking puzzle.

\section*{ACKNOWLEDGMENTS}
We thank A. Bartl, D. Demir, T. Gajdosik, T. Kobayashi, S. Khalil, 
E. Lunghi, H. Murayama, M. Piai, W. Porod, L. Silvestrini and 
H. Stremnitzer as co--authors of some recent works reported in this review. 
We are grateful to G. Barenboim, S. Bertolini and F.J. Botella,  
for enlightening conversations. 
The work of A.M. was partly supported by
the RTN European programs "Across the Energy Frontier" and 
"Early Universe and Supersymmetry" contracts numbers HPRN-CT-2000-0148 and 
HPRN-CT-2000-0152, respectively; O.V. acknowledges financial support from 
a Marie Curie EC grant (HPMF-CT-2000-00457) and partial support from 
spanish CICYT AEN-99/0692.


\end{document}